\documentclass[aps,prb,twocolumn]{revtex4}

\usepackage{graphicx}

\begin{document}
\title{Properties of short channel ballistic carbon nanotube transistors
with ohmic contacts}
\author{Fran\c{c}ois L\'{e}onard$^{\ast }$}
\address{Sandia National Laboratories, Livermore, CA 94551}
\author{Derek A. Stewart$^{\dagger }$}
\address{Cornell Nanoscale Science and Technology Facility, Ithaca, NY 14853}
\date{\today }

\begin{abstract}
We present self-consistent, non-equilibrium Green's function calculations of
the characteristics of short channel carbon nanotube transistors, focusing
on the regime of ballistic transport with ohmic contacts. We first establish
that the band lineup at the contacts is renormalized by charge transfer,
leading to Schottky contacts for small diameter nanotubes and ohmic contacts
for large diameter nanotubes, in agreement with recent experiments. For
short channel ohmic contact devices, source-drain tunneling and
drain-induced barrier lowering significantly impact the current-voltage
characteristics. Furthermore, the ON state conductance shows a temperature
dependence, even in the absence of phonon scattering or Schottky barriers.
This last result also agrees with recently reported experimental
measurements.
\end{abstract}


\maketitle

\section{Introduction}

Since their original fabrication\cite{tans,martel}, carbon nanotube (NT)
transistors have seen much improvement in their performance\cite{fuhrer},
and in the understanding of the science that governs their operation. For
example, it has recently been demonstrated that the metal used to contact
the NT has a strong influence on the device behavior, with reports of ohmic
contacts for Pd\cite{dai1,dai2} and Au\cite{mceuen}, and Schottky contacts
for Ti\cite{ibm1}. Furthermore, it has been demonstrated that scaling of
Schottky barrier NT transistors is very different from traditional
transistors\cite{ibm2}. These studies highlight the danger of using concepts
from conventional silicon devices and applying them to NT devices. In such
circumstances, modeling can play a key role in predicting device behavior,
and in supporting experimental conclusions.

While some theoretical and modeling work beyond traditional transistor
models has been done to study NT transistors, these calculations have used
various approximations, either as simplified treatment of the NT electronic
properties\cite{ibm2}, or by simplifying the calculation of the current at
finite bias voltage\cite{ibm2,leonard}. These approximations prevent a more
detailed and quantitative study of the properties of NT devices,
particularly under non-equilibrium conditions. Here, we present a full
self-consistent calculation that treats the NT electronic structure and the
non-equilibrium quantum electron transport from an atomistic perspective.
This allows the calculation of the charge, electrostatic potential and
current in the device from a quantum transport approach, and includes
quantum effects such as energy quantization and tunneling. It also naturally
captures distortions of the NT bandstructure due to the spatially-varying
electrostatic potential and is amenable to extensions to include scattering
with defects, surfaces, phonons, etc.

The calculations presented here provide an approach for evaluating the
performance of ultimate NT transistors, those with ballistic transport,
ohmic contacts and nanometer channels (a recent paper presented a similar
approach\cite{guo}, but focused on Schottky barrier NT transistors). We show
that side contacts between NTs and metal are governed by charge transfer and
band re-alignment, leading to a NT diameter above which contacts are ohmic,
in agreement with recent experiments\cite{kim,chen}. For the ohmic contact
devices, short channel effects can dominate the device behavior for
relatively large channel to gate length ratios. Furthermore, while the
temperature dependence of the ON state conductance in NT devices can be
dominated by Schottky contacts or phonon scattering, we show that, even in
the absence of these effects, the conductance can have a strong temperature
dependence. In fact, we find that the ON state conductance decreases with
increasing temperature, in contrast to Schottky barrier NT transistors and
in agreement with recent experimental measurements on Pd-contacted NT
transistors\cite{dai1}.

\section{Modeling approach}

As shown in figures~1 and 2, the system consists of an infinitely long
single-wall zigzag NT embedded in metal at its two ends and coated by a
dielectric in the channel region. The gate insulator dielectric constant $%
\varepsilon $ is equal to 3.9, as for SiO$_{2}$, and in this work the
dielectric is wrapped by a cylindrical gate of radius 3 nm in the channel
region (unless otherwise stated). The nanotube sits in the middle of a
smooth cylindrical hole in the metal and dielectric, and we take a spacing
of $s=$0.3 nm between the NT and the metal or dielectric walls. The length
of each contact $\lambda $ is equal to 4.17 nm, sufficiently long to
converge the electrostatic potential in the contacts.

\begin{figure}[h]
\includegraphics[height=180pt,width=240pt]{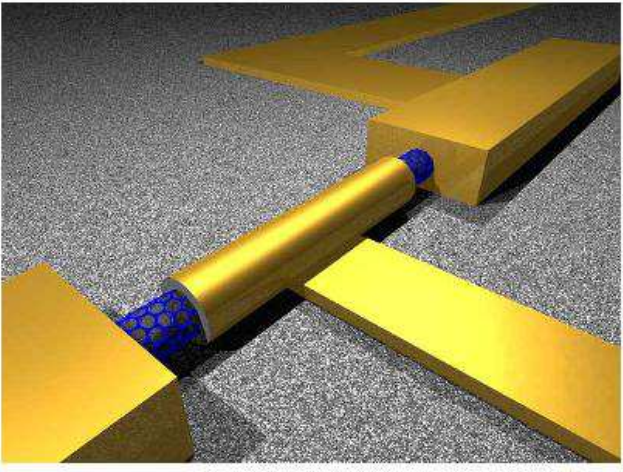}
\caption{Sketch of the NT transistor. The NT is embedded in metals at its
two ends, and in a dielectric in the channel region. The dielectric is
wrapped by a cylindrical gate of 3 nm radius. The separation between the
contacts and the central dielectric region is to illustrate the structure in
the channel; in the calculations, the contacts touch the dielectric.}
\end{figure}

The electronic properties of the NT are described using a tight-binding
framework with one $\pi $ orbital per carbon atom, and a coupling $\gamma
=2.5$ eV\cite{wildoer} between nearest-neighbor atoms. In our calculations,
we assume that the electronic structure of the NT is not perturbed by atomic
interactions with the metal or the dielectric.

To calculate the electronic current flowing through the NT, we apply the
non-equilibrium Green's function formalism\cite{datta} to the NT device,
dividing the NT in principle layers, with each layer corresponding to a ring
of the zigzag NT. The device is divided into a scattering region (layers $1$
to $N$) connecting the lead regions, which consist of semi-infinite
nanotubes embedded in the source and drain contact metals. This approach is
illustrated in Fig. 2.

\begin{figure}[h]
\includegraphics[height=160pt,width=240pt]{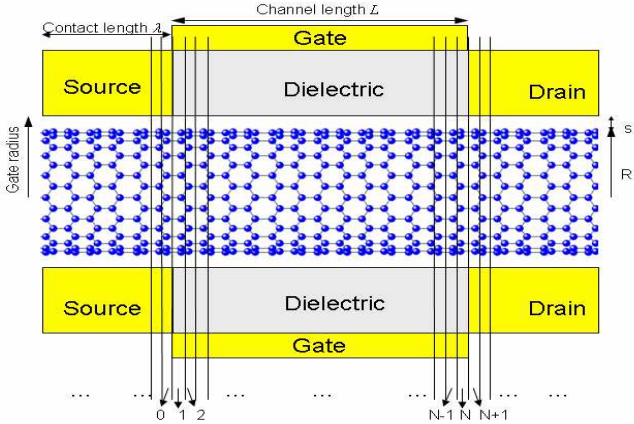}
\caption{Sketch of a cross-section of the nanotube transistor, showing the
important device dimensions and the principle layers used for the quantum
transport calculations.}
\end{figure}

For this geometry, the current can be obtained from

\begin{equation}
I=\frac{4e\gamma }{\pi \hbar }\int dE%
\mathop{\rm Re}%
G_{N,N+1}^{<}
\end{equation}%
where $G^{<}$ is the Green's function, indexed according to the principle
layers in the device, with $N$ total layers in the scattering region. $G^{<}$
is calculated by solving the matrix equations 
\begin{equation}
G^{<}=G^{R}\Sigma ^{<}G^{R\dagger }  \label{g<}
\end{equation}%
and

\begin{equation}
G^{R}=\left[ \left( E-eU\right) I-H_{0}-\Sigma _{L}^{R}-\Sigma _{R}^{R}%
\right] ^{-1}\text{,}  \label{gr}
\end{equation}%
where $H_{0}$ is the tight-binding Hamiltonian for the isolated NT and $U$
is the electrostatic potential on each layer of the system. In our
tight-binding representation, the Hamiltonian matrix elements are $%
H_{0}^{2p,2p-1}=H_{0}^{2p-1,2p}=2\gamma \cos \left( \frac{\pi J}{M}\right) $%
, $H_{0}^{2p,2p+1}=H_{0}^{2p+1,2p}=\gamma $ where $M$ is the number of atoms
around a NT ring and $J=1,...,M$ labels each of the NT bands. Because of the
cylindrical symmetry, we take the electrostatic potential on every atom of a
ring to be the same, allowing substantial reduction in the size of the
matrices (from $MN\times MN$ to $N\times N$). The functions $\Sigma
_{L,R}^{R}$ and $\Sigma ^{<}$ represent the interaction of the scattering
region with the semi-infinite NT left $(L)$ and right $(R)$ leads. These
functions are given by%
\begin{equation}
\Sigma _{L,R}^{R}=\tau _{L,R}^{\dagger }g_{L,R}^{R}\tau _{L,R}
\label{sigmar}
\end{equation}%
and%
\begin{equation}
\Sigma ^{<}=-2i%
\mathop{\rm Im}%
\left( f_{L}\Sigma _{L}^{R}+f_{R}\Sigma _{R}^{R}\right)
\end{equation}%
where $\tau _{L,R}$ is the matrix that couples the scattering region to the
left and right leads, respectively, $g_{L,R}^{R}$ is the Green's function
for the semi-infinite left and right leads and $f_{L,R}$ are the left and
right fermi functions. For the zigzag NT in Fig. 2, and within a
nearest-neighbor tight-binding representation, only the first and last rings
couple with the leads. Therefore, the coupling matrices $\tau _{L,R}$ each
have only one non-zero element, equal to $\gamma $, and we have the non-zero
components%
\begin{eqnarray}
\left[ \Sigma _{L}^{R}\right] _{11} &=&\gamma ^{2}\left[ g_{L}^{R}\right]
_{00}  \label{sigmaR2} \\
\left[ \Sigma _{R}^{R}\right] _{NN} &=&\gamma ^{2}\left[ g_{R}^{R}\right]
_{N+1N+1}  \nonumber
\end{eqnarray}%
where the indices of $g_{L}^{R}$ $\left( g_{R}^{R}\right) $ run from $%
-\infty $ to $0$ $\left( N+1\text{ to }\infty \right) $. The surface Green's
functions $g_{L,R}^{R}$ are calculated using an iterative layer doubling
technique\cite{lopez}, while $G^{R}$ is calculated by using an efficient
approach that iteratively builds up the Green's function at each layer
across the device region\cite{lake}.

The electrostatic potential is calculated by solving Poisson's equation with
a spatially-dependent dielectric constant $\kappa $%
\begin{equation}
\nabla \cdot \left( \kappa \nabla U\right) =-\frac{1}{4\pi \varepsilon _{0}}%
\sigma
\end{equation}
in three-dimensional cylindrical coordinates on a grid with the source
charge on the NT $\sigma $, and with boundary conditions at the source,
drain, and gate surfaces, and at the boundaries of the simulation cell. We
model the interface between the oxide and vacuum with the function $2\kappa
(r)=\left( \varepsilon +1\right) +\left( \varepsilon -1\right) \tanh \left[
\left( r-R-s\right) /\xi \right] $ where $\xi $ is a length scale on the
order of a lattice constant.

In cylindrical coordinates, the above equation is%
\begin{widetext}
\begin{equation}
\frac{\partial ^{2}U\left( r,\phi ,z\right) }{\partial r^{2}}+\frac{1}{r}%
\frac{\partial U\left( r,\phi ,z\right) }{\partial r}+\frac{\partial
^{2}U\left( r,\phi ,z\right) }{\partial z^{2}}+\frac{1}{\kappa (r)}\frac{%
\partial \kappa (r)}{\partial r}\frac{\partial U\left( r,\phi ,z\right) }{%
\partial r}=-\frac{1}{4\pi \varepsilon _{0}\kappa (r)}\sigma (r,\phi ,z)
\end{equation}%
\end{widetext}
with the boundary conditions%
\begin{eqnarray}
U(r &>&R+s,\phi ,z\leq -L/2)=0  \nonumber \\
U(r &>&R+s,\phi ,z\geq L/2)=V_{ds} \\
U(r &=&R_{g},\phi ,-L/2<z<L/2)=V_{gs}  \nonumber \\
\left. \frac{dU(r,\phi ,z)}{dr}\right| _{r=0} &=&0  \nonumber \\
\left. \frac{dU(r,\phi ,z)}{dz}\right| _{z=0} &=&0  \nonumber \\
\left. \frac{dU(r,\phi ,z)}{dz}\right| _{z=L+2\lambda } &=&0  \nonumber
\end{eqnarray}%
where $L$ is the channel length, $V_{ds}$ is the drain-source voltage and $%
V_{gs}$ is the gate-source voltage. As will be discussed below, we use a
uniform distribution of the charge in the azimuthal direction, thus reducing
the three-dimensional partial differential equation to a two-dimensional
partial differential equation in $r$ and $z$. To solve this two-dimensional
equation numerically, we use a finite-difference scheme%
\begin{widetext}
\begin{equation}
\frac{U_{i+1,j}+U_{i-1,j}-2U_{i,j}}{\Delta _{r}^{2}}+\frac{1}{r_{i}}\frac{%
U_{i+1,j}-U_{i-1,j}}{2\Delta _{r}}+\frac{U_{i,j+1}+U_{i,j-1}-2U_{i,j}}{%
\Delta _{z}^{2}}+\frac{1}{\xi }\frac{\varepsilon -1}{\varepsilon }\frac{%
U_{i_{c}+1,j}-U_{i_{c},j}}{\Delta _{r}}=-\frac{1}{4\pi \varepsilon
_{0}\kappa _{i,j}}\sigma _{i,j}  \label{poisson}
\end{equation}%
\end{widetext}
where the subscripts $i,j$ denote a function evaluated at position $\left(
r_{i},z_{j}\right) $ with $r_{i}=(i-1)\Delta _{r}$ and $z_{j}=(j-1)\Delta
_{z},$ and $i_{c}=(R+s)/\Delta _{r}$ is the radial grid point where the
contacts and dielectric begin. The dielectric/vaccum boundary term was
evaluated using a forward difference scheme, assuming that $\xi \ll \Delta
_{r}$; in practice we use a value $\xi =0.0085$ nm and we verified that the
ratio of the electric fields equals the ratio of the dielectric constants at
the interface. The finite-difference boundary conditions are%
\begin{eqnarray}
U_{i\geq i_{c},j\leq j_{S}} &=&0 \\
U_{i\geq i_{c},j\geq j_{D}} &=&V_{ds}  \nonumber \\
U_{i_{G},j_{S}<j<j_{D}} &=&V_{gs}  \nonumber \\
U_{1,j} &=&U_{2,j}  \nonumber \\
U_{i,1} &=&U_{i,2}  \nonumber \\
U_{i,j_{D}+j_{S-1}} &=&U_{i,j_{D}+j_{S}}  \nonumber
\end{eqnarray}%
where $j_{S}=\lambda /\Delta _{z}$ is the axial grid point where the source
contact ends, $j_{D}=\left( L+\lambda \right) /\Delta _{z}$ is the axial
grid point where the drain contact begins, and $i_{G}$ is the radial grid
point at the gate radius. In our calculations we use grid spacings of $%
\Delta _{z}=$0.07 nm in the axial direction and $\Delta _{r}=$0.03 nm in the
radial direction. Numerical solution of Eq. $\left( \ref{poisson}\right) $
is obtained using a successive over-relaxation procedure\cite{recipes}. Once
the electrostatic potential is obtained, the value for $U$ on each layer of
the system (i.e. that enters the Hamiltonian) is taken as the value of the
electrostatic potential at the atomic position of each ring along the NT.

To obtain the charge density, we note that the tight-binding technique only
provides the total charge on each layer; in our formalism the total charge
on layer $l$ is given by $\frac{e}{2\pi }\int dE%
\mathop{\rm Im}%
G_{ll}^{<}$, which needs to be spatially distributed. We assume a uniform
distribution of the charge in the azimuthal direction, and spatially
distribute the total charge in the radial and axial directions with a
Gaussian smearing function. Thus the three-dimensional charge density is
given by 
\begin{equation}
\sigma (r,\phi ,z)=\sum_{l}g(z-z_{l},r-R)\frac{e}{2\pi }\int dE%
\mathop{\rm Im}%
G_{ll}^{<}  \label{charge}
\end{equation}%
where $g(z-z_{l},r-R)=\left( 4\pi ^{2}R\sigma _{z}\sigma _{r}\right)
^{-1}\\ \exp \left[ -(z-z_{l})^{2}/2\sigma _{z}^{2}\right] \exp \left[
-(r-R)^{2}/2\sigma _{r}^{2}\right] $ with $R$ the tube radius, $z_{l}$ the
position of ring $l$, and $\sigma _{z}$ and $\sigma _{r}$ the smearing
lengths in the axial and radial directions respectively (this expression for 
$g$ is valid when $R\gg \sigma _{r}$, and we use values of $\sigma _{z}=0.14$
nm and $\sigma _{r}=0.06$ nm).

Our overall procedure is to start from some initial guess for the potential
and solve for the Green's functions from the system of Eqs $\left( \ref{g<}-%
\ref{sigmaR2}\right) $, calculate the charge distribution in the device from
Eq. $\ref{charge}$, and then re-calculate the potential profile across the
device based on solving Poisson's equation. This process is iterated until
both the potential and charge distribution converge; in practice this is
done using a simple mixing of the charge at each iteration step.

As discussed in a previous work\cite{leonard}, short channel NT transistors
can possess two different operating regimes: a conventional transistor
regime, and a Coulomb blockade regime at large gate voltages due to channel
inversion. While simplified models exist to describe the Coulomb blockade
regime\cite{leonard}, a proper treatment involves including the
electron-electron interaction self-consistently in the Green's function
approach. In this paper, we focus on the transistor regime and leave the
Coulomb blockade regime for future work.

\section{Properties of contacts}

Before discussing the full current-voltage characteristics of the NT
transistor, we begin by discussing the properties of the contacts for NTs
embedded in metals. Fig. 3 shows a cross-section of the contact, and the
associated bare band lineup. For such a contact, the difference between the
metal Fermi level $E_{F}$ and the valence band edge $E_{v}$ before charge
transfer is simply given by%
\begin{equation}
E_{F}-E_{v}=\Phi _{NT}-\Phi _{m}+\frac{1}{2}E_{g}  \label{bare}
\end{equation}%
where $\Phi _{m}$ and $\Phi _{NT}$ are the metal and NT workfunctions
respectively, and $E_{g}$ is the NT band gap. A positive value for $%
E_{F}-E_{v}$ indicates a Schottky barrier, while a negative value indicates
an ohmic contact (here we focus on the barrier for holes, the relevant
quantity for high workfunction metals used in most experiments; the barrier
for electrons is $E_{v}+E_{g}-E_{F}$ and would be relevant for low
workfunction metals. Because the NT band structure is symmetric around the
midgap, the results presented below apply to both cases). Given the relation 
$E_{g}=a\gamma /d$ between bandgap and NT diameter $d$ ($a=0.142$ nm is the
C-C bond length), the band lineup thus depends on the NT diameter. The
behavior of Eq. \ref{bare} is shown in Fig. 4 for two different values of $%
\Phi _{m}-\Phi _{NT}$ as a function of the NT diameter. This simple picture
for the band lineup is modified due to charge transfer between the metal and
NT, as we now discuss.

\begin{figure}[h]
\includegraphics[height=125pt,width=240pt]{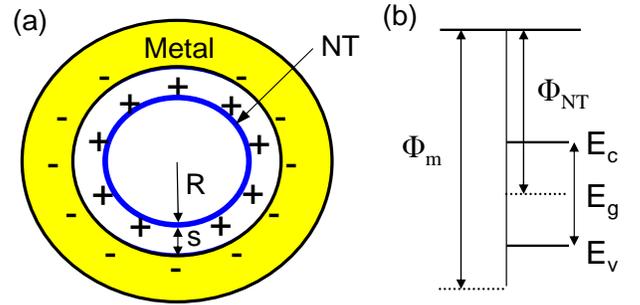}
\caption{Panel (a) shows a sketch of the cross-section of the nanotube/metal
contact, while panel (b) is the bare band lineup.}
\end{figure}

Because the metal Fermi level is not in line with the charge neutrality
level in the NT (midgap for an undoped NT), charge transfer between the
metal and NT occurs, as illustrated in Fig. 3a. From an image potential
argument, this transferred charge creates a charge dipole at the
nanotube/metal interface, and an associated electrostatic potential. This
electrostatic potential in turn shifts the bands, and changes the amount of
transferred charge. Thus, the charge and potential must be determined
self-consistently. This behavior can be described using simple models for
the charge and the potential (the results of the atomistic approach
presented in the previous section will be discussed further below). The
charge per atom on the NT can be expressed as%
\begin{equation}
\sigma =eN\int D(E)f(E-E_{F})dE  \label{sigma}
\end{equation}%
where%
\begin{equation}
D(E)=\frac{a\sqrt{3}}{\pi ^{2}R\gamma }\frac{\left| E+eV_{NT}\right| }{\sqrt{%
\left( E+eV_{NT}\right) ^{2}-\left( E_{g}/2\right) ^{2}}}
\end{equation}%
is the density of states, $f\left( E-E_{F}\right) $ is the Fermi function
and $N=4/(3\sqrt{3}a^{2})$ is the atomic areal density. We assume a uniform
distribution of the charge on the NT.

\begin{figure}[h]
\includegraphics[height=340pt,width=240pt]{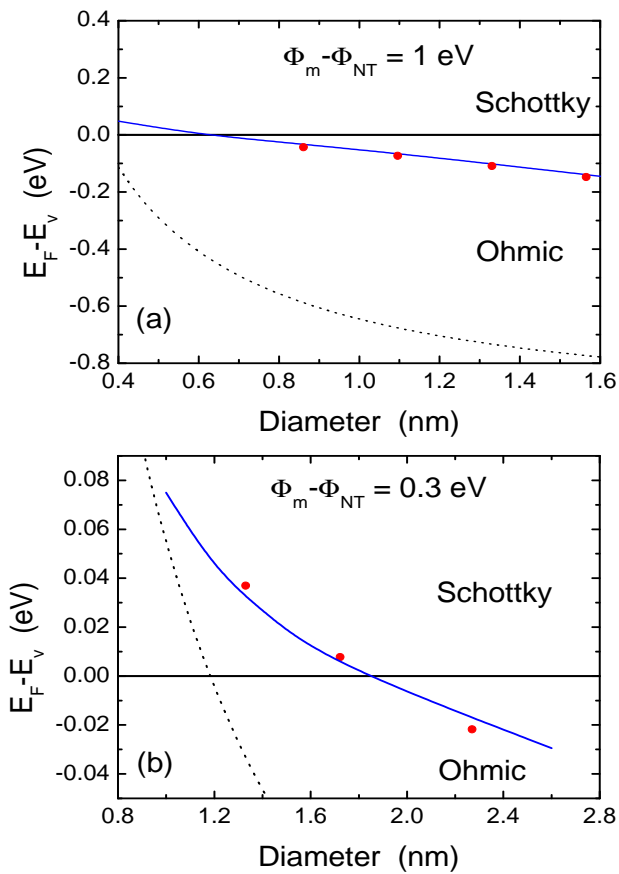}
\caption{Band lineup at nanotube/metal contacts. Dotted line is prediction
from Eq. (\ref{bare}), solid line is computed from Eqs (\ref{sigma}) and (%
\ref{potential}), and circles are calculated from atomistic approach.}
\end{figure}

For the geometry of Fig. 3a, solution of Poisson's equation gives the
potential on the NT as%
\begin{equation}
eV_{NT}=-\sigma \frac{eR}{\varepsilon _{0}}\ln \frac{R+s}{R}.
\label{potential}
\end{equation}%
For a given NT, equations \ref{sigma} and \ref{potential} can be solved
self-consistently. The solid lines in Fig. 4 show results of such
calculations for the room temperature alignment between the metal Fermi
level and the NT band edge for $\Phi _{m}-\Phi _{NT}=1$ eV and $0.3$ eV.
Comparison with the barrier predicted from Eq. \ref{bare} (dotted lines in
the figure) indicates that the charge transfer significantly changes the
band line-up. While for the higher workfunction difference the contacts are
ohmic down to very small diameter NTs (Fig. 4a), for the lower workfunction
metal (Fig. 4b), the results indicate a transition between Schottky and
ohmic behavior at a NT diameter around 1.8 nm. This result agrees with
recently published experimental data\cite{kim,chen}.

The results of this simplified analytical model can be compared with
calculations from the atomistic, self-consistent, Green's function approach
introduced in the previous section. For this comparison, the electrostatic
potential on the NT from the atomistic approach is taken to be $U_{i_{R},1}$
where $i_{R}=R/\Delta _{r}$ is the grid point at the NT radius and we use $%
V_{sd}=V_{gs}=0$ (since the contact length is long enough to screen the
electric fields from the drain or gate electrodes, the value of $U_{i_{R},1}$
is independent of the choice of $V_{ds}$ and $V_{gs}$). The data points in
Figure 4 show results of such calculations for several semiconducting zigzag
NTs, demonstrating excellent agreement with the simplified approach
presented above.

\section{Transistor characteristics}

Having established the conditions needed to make ohmic contacts to NTs, we
now discuss the room-temperature behavior of NT transistors with such
contacts. For explicit calculations, we take the NT midgap as the energy
reference level, take $\Phi _{m}-\Phi _{NT}=$ 1 eV, and use a (17,0) zigzag
NT. This NT has a radius of 0.66 nm and in our tight-binding description, it
has a bandgap of 0.55 eV.

The current versus gate-source voltage characteristics of the NT transistor
calculated using the atomistic approach of section II are shown in Fig. 5
for two different values of the channel length, and for source-drain
voltages of -0.1 and -0.3 V. The general behavior consists of an ON state at
smaller values of $V_{gs}$ and an OFF state at larger values of $V_{gs}$.
These two regimes originate from the presence or absence of a barrier
blocking the hole current, as shown in the band diagrams of Fig. 5b. The 10
nm channel device shows rather poor characteristics, with an ON/OFF ratio of
only 1000 and a subthreshold swing $S=\left( d\log I/dV_{gs}\right)
^{-1}=160 $ mV/decade. This behavior originates from poor control of the
gate over the electrostatics in the channel, and from tunneling across the
hole barrier in the OFF regime. Increasing the channel length to 20 nm as in
Fig. 5b gives a much better $S=69$ mV/decade and improved ON/OFF ratio. A
long-channel transistor with undoped channel is expected to be maximally OFF
at $V_{gs}=0$; it is clear from Fig. 5 that even the 20 nm device is ON at
that gate voltage. The inset in Fig. 5b shows the calculated band bending
for $V_{gs}=0 $ and $V_{ds}=0$, indicating that the channel is effectively
doped $p$-type due to a long-ranged charge transfer from the contacts\cite%
{leonard-pn}. Thus the NT device is normally ON, despite the fact that the
channel is undoped.

\begin{figure}[h]
\includegraphics[height=340pt,width=240pt]{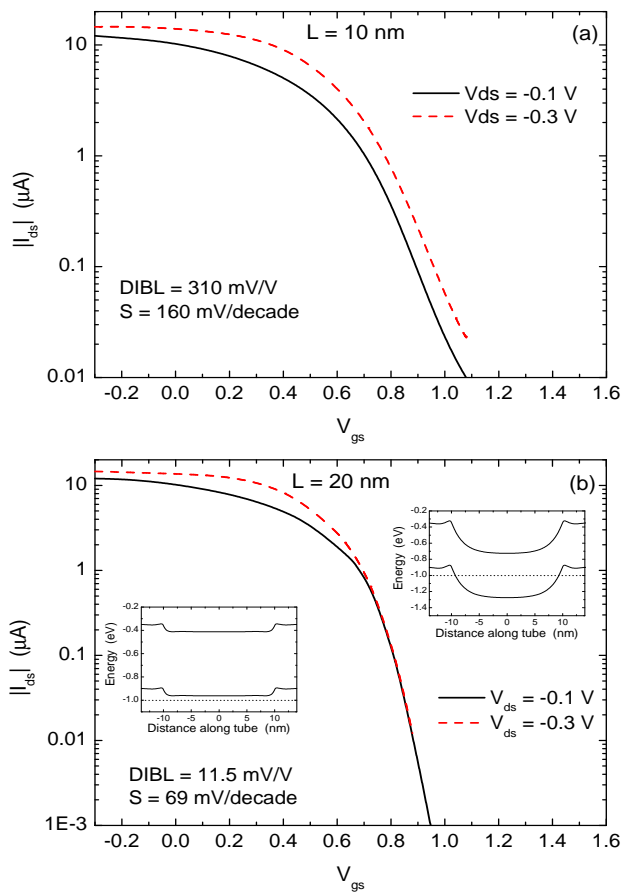}
\caption{Current as a function of gate-source voltage for channel lengths of
(a) 10 nm and (b) 20 nm. The insets show the band bending for $V_{gs}=0$
(left) and $V_{gs}=1$ V (right).}
\end{figure}

The output characteristics of the NT transistor are shown in Fig.~6,
indicating good saturation of the current for the 20 nm channel device, but
no saturation for the 10 nm device. This behavior can be attributed to
drain-induced barrier lowering (DIBL) as indicated in Fig. 7. There, it is
shown that for the 10 nm channel device, application of a source-drain
voltage reduces the barrier between the source Fermi level and the middle of
the channel. A signature of the DIBL effect is also seen in Fig. 5a, where
there is significant shift between the transfer characteristics for
source-drain voltages of $-0.1$ and $-0.3$ Volts. A quantitative calculation
of the DIBL from $\Delta I/\Delta V_{ds}$ at 0.1 $\mu A$ gives a very large
value of 310 mV/V. Fig. 5b shows that the DIBL for the 20 nm channel length
is reduced substantially to 11.5 mV/V, consistent with the saturation of the
output characteristics of Fig. 6.

\begin{figure}[h]
\includegraphics[height=320pt,width=240pt]{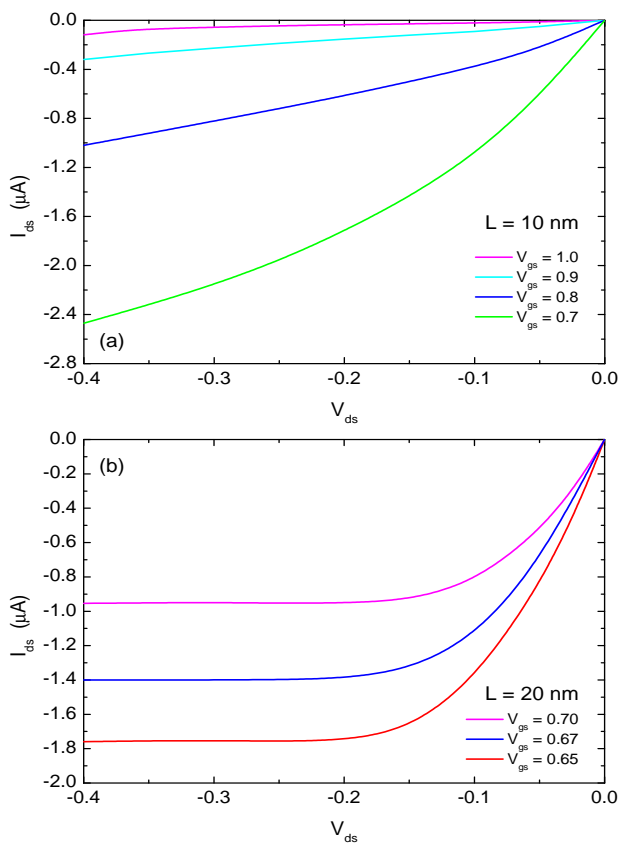}
\caption{Current as a function of drain-source voltage for (a) 10 nm channel
length and (b) 20 nm channel length.}
\end{figure}

The reduction of the subthreshold swing and the DIBL with increasing channel
length is captured in Fig. 8. It is clear that both quantities vary rapidly
over the same length scale of a few nanometers. Typical device constraints
require DIBL less than 100 mV/V and $S$ less than 80 mV/dec; this restricts
the nanotube transistor in this geometry and with this dielectric constant
to a channel length greater than 15 nanometers, outside of the shaded areas
in the figure. To extend this analysis, we repeated the calculations of the
subthreshold swing for gate radii of 1.5, 6 and 9 nm; the inset in Fig. 8a
shows that a good scaling behavior can be obtained if the channel length is
scaled by a quantity $\alpha \sim \sqrt{R_{g}+l}$, where $l=1$ nm. While we
have not been able to derive an analytical expression to justify this
scaling behavior, we note that electrostatic analyses for cylindrical gate
transistors\cite{oh,john} predict a scaling quantity proportional to the
oxide thickness $\alpha \sim \left( R_{g}-R-s\right) $; we have tried this
type of relation and find that it does collapse the data onto a single
curve. We suspect that the unusual dielectric response of the NT\cite%
{screening}, strong charge transfer from the contacts and the actual device
geometry render the conventional analyses inapplicable; more work is needed
to address these issues.

\begin{figure}[h]
\includegraphics[height=380pt,width=240pt]{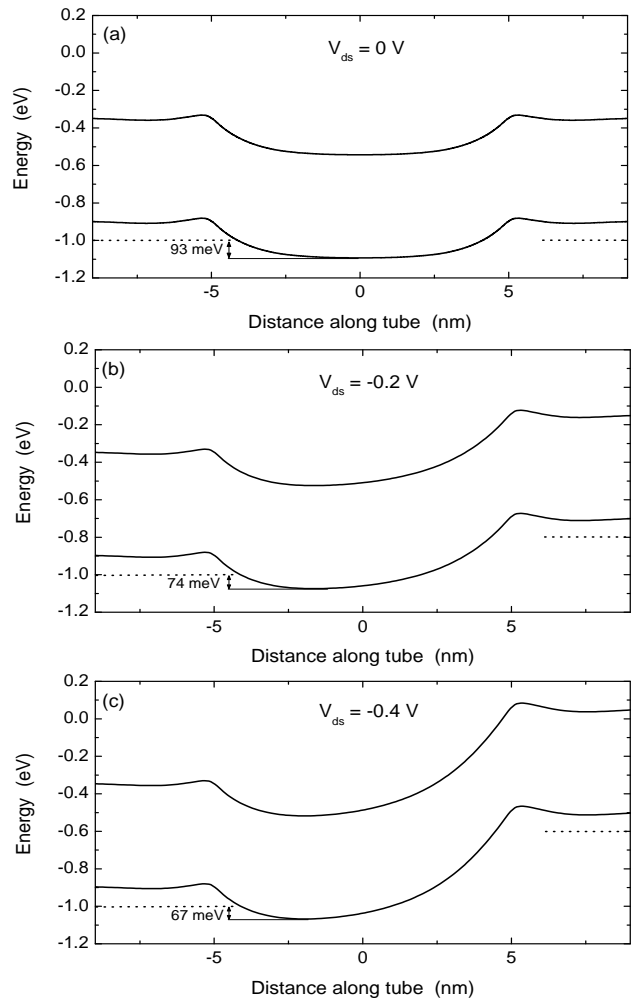}
\caption{Calculated self-consistent band bending for the 10 nm channel
device for a gate-source voltage of 0.8 V. The solid lines are the valence
and conduction band edges; horizontal dotted lines are the metal Fermi
levels.}
\end{figure}

\begin{figure}[h]
\includegraphics[height=330pt,width=240pt]{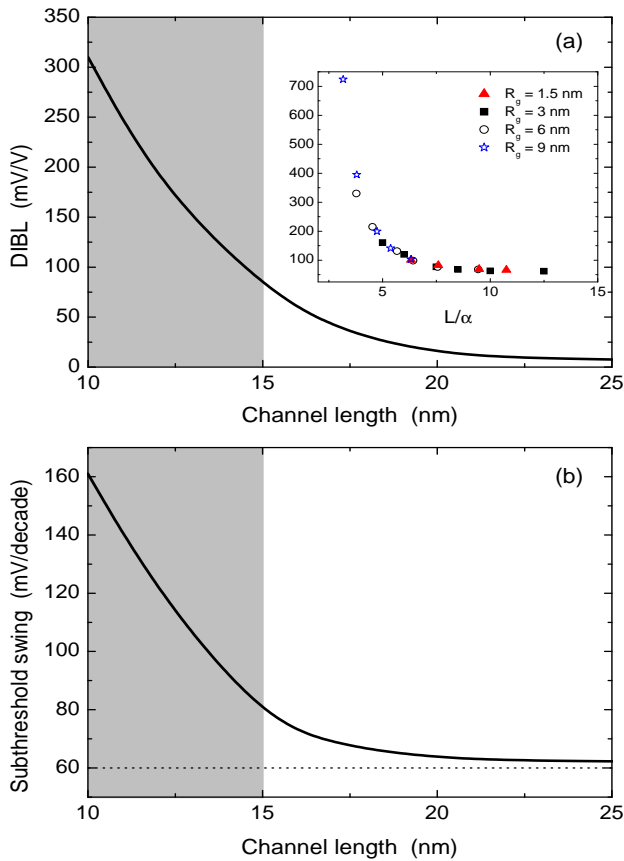}
\caption{Panels (a) and (b) show the variation of DIBL and subthreshold
swing on channel length, respectively. Shaded areas are regions where the
short channel effects are larger than typical device requirements. The inset
in panel (a) shows collapse of the data for several devices of different
channel lengths and gate radii upon scaling of the channel length. The
horizontal dotted line in panel (b) is the theoretical limit for the
subthreshold swing.}
\end{figure}

\section{Temperature dependence of ON state conductance}

As we discussed above, one of the important issues for carbon nanotubes is
the nature of contacts between NTs and metals\cite{kim,Fermipinning,chen}. A
useful way to elucidate the properties of contacts is to study the
temperature dependence of the device characteristics\cite{dai1,appenzeller}.
In NT devices, temperature is useful to study the role of phonon scattering
and Schottky barriers, as both of these lead to a temperature dependence of
the conductance. We now show that even for the phonon-free, ballistic-ohmic
NT device presented here, the conductance can have a strong temperature
dependence. Figure 9 shows the calculated ON state conductance from our
non-equilibrium Green's function approach as a function of temperature for $%
\Phi _{m}-\Phi _{NT}=$ 1 eV and $\Phi _{m}-\Phi _{NT}=$ 0.6 eV. Clearly, the
conductance decreases with increasing temperature, with a stronger effect
for the smaller metal workfunction. To understand the temperature dependence
of the conductance $C$, we note that in the ON state the bands are flat with
the Fermi level in the valence band, and we consider the expression\cite%
{datta}%
\begin{equation}
C=\frac{4e^{2}}{h}\int_{-\infty }^{\infty }P(E)\left( -\frac{\partial f}{%
\partial E}\right) dE  \label{cond1}
\end{equation}%
where $P$ is the transmission probability and $f$ is the Fermi function.
Assuming that $P(E)=1$ for $E<E_{v}=E_{F}+\Delta $, we obtain

\begin{equation}
C=\frac{4e^{2}}{h}\frac{e^{\Delta /kT}}{1+e^{\Delta /kT}}.  \label{cond2}
\end{equation}%
Thus, the conductance is sensitive to the self-consistent position of the
metal Fermi level with respect to the valence band edge, $\Delta
=E_{v}-E_{F} $. In principle, the value of $\Delta $ depends on the contact
capacitance and temperature, and requires numerical evaluation at each
temperature. But to illustrate the behavior, the dotted lines in Fig.~9 show
the results of Eq.~$\left( \ref{cond2}\right) $ for a constant $\Delta $
equal to its zero temperature value (we scaled the expression to fit the
zero temperature conductance calculated numerically). Equation $\left( \ref%
{cond2}\right) $ indicates that the conductance can vary between the zero
temperature limit of $4e^{2}/h$ and the high temperature limit of $2e^{2}/h$%
, with the variation occurring over a temperature range $kT\sim \Delta $.
Higher workfunction metals should thus show a weaker dependence on
temperature. Indeed, the metal with $\Phi _{m}-\Phi _{NT}=$ 1 eV gives $%
\Delta \approx 0.1 $ eV and its conductance varies more slowly than the
metal with $\Phi _{m}-\Phi _{NT}=$ 0.6 eV that has $\Delta \approx 0.02$ eV.
The important point however is that the device conductance can vary with
temperature by a factor of 2 even though there is no phonon scattering or
Schottky barriers. Figure 9 also shows that recent experimental data\cite%
{dai1} for NT devices with Pd contacts agree well with the predicted
behavior for $\Delta =0.02$ eV.

\begin{figure}[h]
\includegraphics[height=180pt,width=240pt]{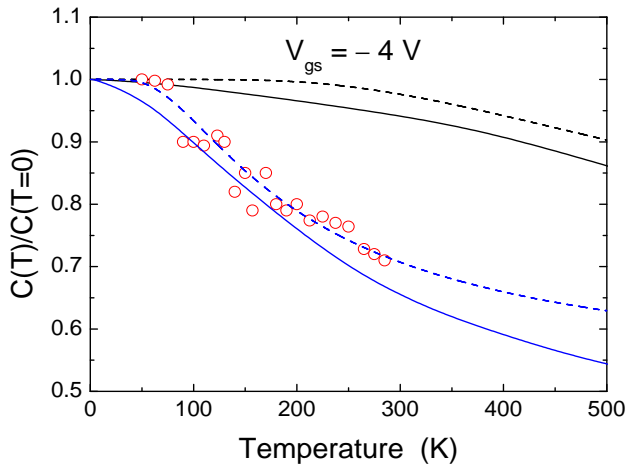}
\caption{Conductance of the nanotube transistor in the ON regime as a
function of temperature. Black and blue solid lines are results of the
atomistic calculations for $\Phi _{m}-\Phi _{NT}=$ 1 eV and 0.6 eV,
respectively. Dotted lines are calculated from Eq.~$\left( \ref{cond2}%
\right) $, see text for details. The circles are experimental data points
extracted from Ref. \protect\cite{dai1}.}
\end{figure}

\section{Conclusion}

There are several main conclusions from this work. First, for nanotubes
embedded in metals, we have shown that charge transfer between the metal and
the NT can renormalize the band lineup, and typically locates the Fermi
level within a few tens of meV from the valence band edge. Second, our
non-equilibrium quantum transport calculations indicate that nano-channel NT
transistors can be dominated by short channel effects due to source-drain
tunneling and drain-induced barrier lowering, even at channel lengths where
Si transistors show good characteristics. Finally, for ohmic contacts, the
ON state conductance decreases with increasing temperature, in contrast to
Schottky barrier NT transistors.

Sandia is a multiprogram laboratory operated by Sandia Corporation, a
Lockheed Martin Company, for the United States Department of Energy under
contract DE-AC01-94-AL85000. The Cornell Nanoscale Science and Technology
Facility is part of the National Nanotechnology Infrastructure Network
(NNIN) funded by the National Science Foundation.

$^{\ast }$E-mail: fleonar@sandia.gov

$^{\dagger }$E-mail: stewart@cnf.cornell.edu


\begin{references}
\bibitem{tans} S. J. Tans, A. R. M. Verschueren, C. Dekker, Nature (London) 
{\bf 393}, 49 (1998).

\bibitem{martel} R. Martel{\it ,} T. Schmidt, H. R. Shea, T. Hertel, Ph.
Avouris, Appl. Phys. Lett. {\bf 73}, 2447 (1998).

\bibitem{fuhrer} B. M. Kim, T. Brintlinger, E. Cobas, M. S. Fuhrer, H.
Zheng, Z. Yu, R. Droopad, J. Ramdani, K. Eisenberg, Appl. Phys. Lett. {\bf 84%
}, 1946 (1998).

\bibitem{dai1} A. Javey, J. Guo, Q. Wang, M. Lundstrom, H. J. Dai, Nature
(London) {\bf 424}, 654 (2003).

\bibitem{dai2} A. Javey, J. Guo, D. B. Former, Q. Wang, D. W. Wang, R. G.
Gordon, M. Lundstrom, H. J. Dai, Nano Lett. {\bf 4}, 447 (2004).

\bibitem{mceuen} Y. Yaish, J.-Y. Park, S. Rosenblatt, V. Sazonova, M. Brink,
P. L. McEuen, Phys. Rev. Lett. {\bf 92}, 046401 (2004).

\bibitem{ibm1} S. Heinze, J. Tersoff, R. Martel, V. Derycke, J. Appenzeller,
Ph. Avouris, Phys. Rev. Lett. {\bf 89}, 106801 (2002).

\bibitem{ibm2} S. Heinze, M. Radosavljevic, J. Tersoff, Ph. Avouris{\it ,}
Phys. Rev. B {\bf 68}, 235418 (2003).

\bibitem{leonard} F. L\'{e}onard and J. Tersoff, Phys. Rev. Lett. {\bf 88},
258302 (2002).

\bibitem{guo} J. Guo, S. Datta, M. Lundstrom, IEEE Trans. Electron Devices 
{\bf 51}, 172 (2004).

\bibitem{kim} W. Kim, A. Javey, R. Tu, J. Cao, Q. Wang, H. Dai, Appl. Phys.
Lett. {\bf 87}, 173101 (2005).

\bibitem{chen} Z. Chen, J. Appenzeller, J. Knoch, Y.-M. Lin, Ph. Avouris,
Nano Lett. {\bf 5}, 1497 (2005).

\bibitem{wildoer} J. W. G. Wild\"{o}er, L. C. Venema, A. G. Rinzler, R. E.
Smalley, C. Dekker, Nature (London) {\bf 391}, 59 (1998); T.W. Odom, J.
Huang, P. Kim, C. M. Lieber, Nature (London) {\bf 391}, 62 (1998).

\bibitem{datta} S. Datta, {\it Electronic transport in mesoscopic systems }%
(Cambridge University Press, Cambridge, England 1995).

\bibitem{lopez} M. P. Lopez-Sancho, J. M. Lopez-Sancho, J. Rubio, J. Phys.
F. Met. Phys. {\bf 15}, 851 (1985).

\bibitem{lake} R. Lake, G. Klimeck, R. C. Bowen, D. Jovanovic, J. Appl.
Phys. {\bf 81}, 7845 (1997).

\bibitem{recipes} W. H. Press {\it et al}, {\it Numerical recipes in fortran
77: The art of scientific computing, 2nd ed. }(Cambridge University Press,
Cambridge, England 1992).

\bibitem{leonard-pn} F. L\'{e}onard and J. Tersoff, Phys. Rev. Lett. {\bf 83}%
, 5174 (1999).

\bibitem{oh} S.-H. Oh, D. Monroe, J. M. Hergenrother, IEEE Electron Device
Lett. {\bf 21}, 445 (2000).

\bibitem{john} D. L. John, C. Castro, D. L. Pulfrey, IEEE Trans. Nano. {\bf 2%
}, 175 (2003).

\bibitem{screening} F. L\'{e}onard and J. Tersoff, Appl. Phys. Lett. {\bf 81}%
, 4835 (2002).

\bibitem{Fermipinning} F. L\'{e}onard and J. Tersoff, Phys. Rev. Lett. {\bf %
84}, 4693 (2000).

\bibitem{appenzeller} J. Appenzeller, M. Radosavljevic, J. Knoch, Ph.
Avouris, Phys. Rev. Lett. {\bf 92}, 048301 (2004).
\end{references}
\end{document}